# The Yield Volume Fraction approach to the description of the stress-strain curve of a nickel-base superalloy

Jingwei Chen, Alexander M. Korsunsky *

*MBLEM, Department of Engineering Science, University of Oxford, Parks Road, Oxford OX1 3PJ, United Kingdom*

jingwei.chen@eng.ox.ac.uk

Alexander.korsunsky@eng.ox.ac.uk, *corresponding author

## Abstract

The stress-strain curves of most metallic alloys are often described using the relatively simple Ramberg-Osgood relationship. Whilst this description captures the overall stress-strain curve under monotonic tensile loading with reasonable overall accuracy, it often presents significant errors in the immediate post-yield region where the interplay between the elastic and plastic strains is particularly significant. This study proposes and develops a new approach to the description of the tensile stress-strain curve based on the Yield Volume Fraction (YVF) function. The YVF description provides an excellent match to experimental stress-strain curves based on a physically meaningful parameter that corresponds to the cumulative volume fraction of the polycrystal that undergoes yielding during monotonic deformation. The statistical nature of the polycrystal yield phenomenon is highlighted by the fact that the YVF model achieves good agreement with observations when the lognormal and extreme value distributions are employed to express the cumulative density function for the total yield volume fraction, and the probability density function for the incremental yield volume fraction, respectively. This proposed approach is compared with crystal plasticity finite element (CPFE) simulations and the Ramberg-Osgood model, along with experimental observations. The results highlight the potential of more extensive use of statistical methods in the description of material deformation response for improved design.

# 1. Introduction

Stress-strain curves are a key representation of the material deformation response that allows revealing the relationship between the material structure and its mechanical properties. Important materials properties such as Young's modulus, yield stress and ultimate tensile strength can be determined, along with the detailed hardening response to plastic deformation. The stress-strain relationships are often represented by fitting suitable mathematical equations to experimental stress-strain data. One of the most popular equations for metallic materials was proposed by Ramberg and Osgood (1943) and modified by Hill (1944). The classic Ramberg-Osgood (R-O) relates the engineering strain and stress by the summation of a linear elastic term and a power-law plastic term

$$\varepsilon = \varepsilon_{el} + \varepsilon_{pl} = \frac{\sigma}{E} + \left(\frac{\sigma}{k}\right)^n \qquad (1)$$

Here $\varepsilon_{el}$ and $\varepsilon_{pl}$ are the elastic and plastic strain, respectively. $E, k$ and $n$ are the elastic modulus, proof stress related parameter, and strain hardening exponent which describes the sharpness of the 'knee' of the stress-strain curve, respectively.

Despite its popularity, the R-O equation is inadequate to represent the full-range stress-strain curves for many alloys. It is well accepted that the Ramberg-Osgood relation can provide an excellent agreement with experimental data for stress below the yield point $\sigma_{0.2}$. For higher stress levels above $\sigma_{0.2}$, R-O predictions may become significantly inaccurate and overestimate the stress (Fernando et al., 2020). Over the past years, many modified models have been developed based on the classic Ramberg-Osgood expression to account for inaccuracy in stress ranges exceeding $\sigma_{0.2}$ (Macdonald et al., 2000; Mirambell & Real., 2000; Rasmussen, 2003; Quach et al.,

2008; Hradil et al., 2013; Dundu, 2018; Fernando et al., 2020). Most of the modified models employ a two-stage or three-stage method to describe the stress-strain curves: one expression for the first stage using Equation 1 up to $\sigma_{0.2}$, and one or two separate equations for the stress level above $\sigma_{0.2}$. In the current research, we will examine the limitations of R-O type relations by placing the consideration in a context of yield statistics.

The present article introduces a simple yet powerful approach to the analysis of stress-strain curves using a statistical method based on the Yield Volume Fraction concept. This treatment gives rise to an integrated method that can match well the stress-strain curves of a nickel-base superalloy up to the strain level of 1.55%. In this approach, the stress-strain curves are related to the cumulative yield volume fraction using lognormal distribution and Gumbel distribution. The results obtained from the proposed method are compared with classical R-O description, CPFE simulation and experimental data.

## 2. Methodology

### 2.1 Experimental stress-strain curve

To develop the proposed method in the present study, the tensile stress-strain curve for a nickel-base superalloy nominated Haynes 282 is employed (Jaladurgam et al., 2020). The experimental curve is illustrated in Figure 1. A total strain ranges up to 1.55% was chosen in this research. We are particularly interested in the regime when the contributions from elastic and plastic strain are of approximately similar magnitude. This represents the most difficult condition for description compared to cases when either elastic or plastic deformations dominate.

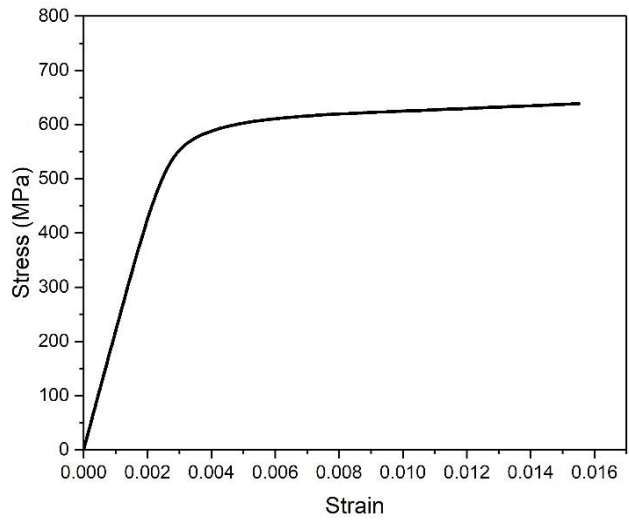

**Figure 1**. The tensile stress-strain curve obtained from the experiment in the literature (Jaladurgam et al., 2020).

## 2.2 CPFE simulation

The CPFE model was implemented using phenomenological crystal plasticity constitutive equations developed by Manonukul & Dunne (2004). It was further extended to account for elastic anisotropy and to allow three-dimensional modelling for alloys with various crystal structures (Song et al., 2008; Korsunsky et al., 2007). Dini et al. (2006; 2009) introduced the diffraction post-processing interpretation to simulate the intragranular strains that can be directly comparable with the results of diffraction measurements. To calibrate the material parameters in CPFE, the overall response of a representative volume element (RVE) with 1500 grains during tensile loading was fitted to the experimental stress-strain curve for Haynes 282. Periodic microstructure and periodic boundary were employed in the RVE shown in Figure 2 (a). Optimal values of material parameters were determined by fitting the macroscopic and mesoscopic response of RVE to the experimental stress-strain curve and neutron diffraction measurement obtained by Jaladurgam et al. (2020). For a detailed description of the constitutive equations and the calibration procedure, readers can refer to our previous published paper (Chen et al., 2022). Figure 2 compares the predicted stress-strain curve with the experiment curve and shows that excellent agreement was achieved.

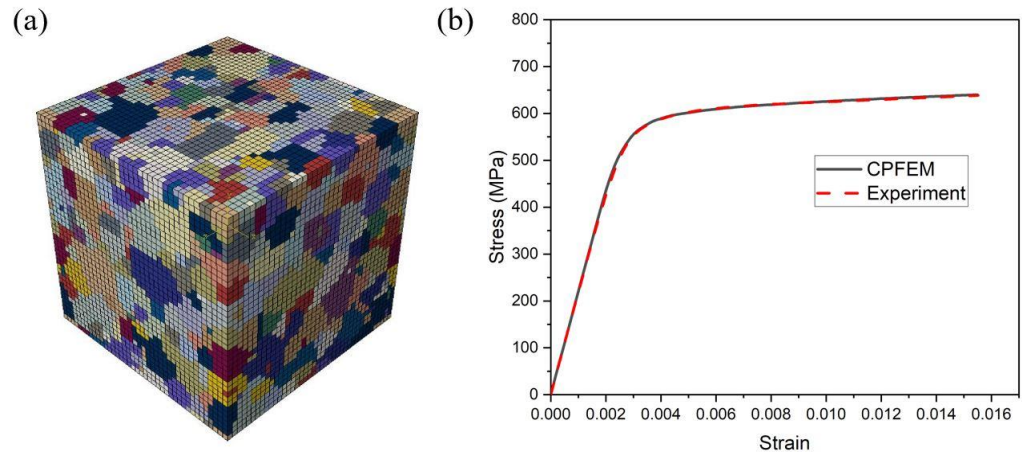

**Figure 2**. (a) The microstructure of the RVE used to calibrate the materials parameters (different colours indicate different grains). (b) The tensile stress-strain curve obtained from CPFE prediction and experimental data.

**2.3 Yield Volume Fraction integrated method**

Most of the metallic materials in use are polygranular, which means that they consist of distinct domains of materials (grains), in terms of crystal orientation, phase composition and lattice structure, etc. In polycrystalline materials such as nickel-base superalloy, both the elastic and plastic deformation are highly inhomogeneous. This occurs due to the heterogeneous nature of polycrystalline material caused by the orientation dependence of elastic and plastic properties. As a numerical measure of the extent of disparity between grains of different orientation, the elastic Zener anisotropy factor $A = 2C_{44}/(C_{11} - C_{12})$ can be used that typically takes values close to 2.7 for nickel-base alloys (with unity corresponding to isotropic material)(Everaerts et al., 2019). As far as yield properties are concerned, the ratio of Schmid factors can be used between the most 'hard' and 'soft' crystal orientations, that shows a milder degree of anisotropy that amounts to about 3% (Ying et at., 2018).

Depending on the orientation of the underlying lattice structure with respect to the loading axis (or, more precisely, the local principal stress axes) and also depending on the elastic anisotropy and neighbouring grains, different grains deform to different extent as a function of the remote applied stress. The onset of yield with polycrystalline materials is a progressive process and it occurs nonlinearly with the increase of macroscopic strain and stress. The progressive yielding is one of the principal causes of macroscopically observed continuous strain hardening in the 'knee' region of the stress-strain curve. Following the concept proposed by Korsunsky (2017) and illustrated in Figure 3(a), we introduce here the cumulative Yield Volume Fraction $\beta(\varepsilon)$ under tensile strain $\varepsilon$. Considering the onset and evolution of plasticity within a polycrystal, it may be expected that the YVF function $\beta(\varepsilon)$ be described by a sigmoid curve. When the RVE contains a sufficient number of grains, the $\beta(\varepsilon) - \varepsilon$ curve will be continuous. The strain levels $\varepsilon_1$ and $\varepsilon_2$ represent the starting and ending stages of the progressive yield process, respectively. The polycrystal will behave purely elastically and no yield occurs when the total strain is less than $\varepsilon_1$. Increasing volume of material yields during further loading until the polycrystal is fully yielded at strain $\varepsilon_2$ (this parameter may be assumed to be very large without any loss of generality).

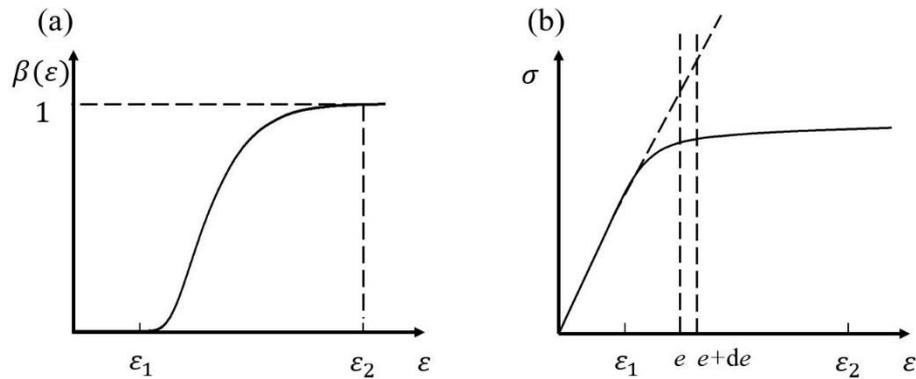

**Figure 3.** (a) Illustration for the cumulative yield volume fraction $\beta(\varepsilon)$ versus strain. (b) The illustration of a stress-strain curve derived from $\beta(\varepsilon)$ (Korsunsky, 2017).

The Yield Volume Fraction description is based on the following relationship between the function $\beta(\varepsilon)$ and the polycrystal stress-strain curve. For simplicity, we shall assume the polycrystal has a sufficient number of grains and has the same Young's modulus $E$, i.e. no elastic anisotropy when we consider the polycrystal as a whole. The stress-strain curve function $\sigma(\varepsilon)$ can be expressed as

$$\sigma(\varepsilon) = \begin{cases} E\varepsilon, & \varepsilon < \varepsilon_1 \\ \sigma_y + \sigma_h, & \varepsilon \geq \varepsilon_1 \end{cases} \qquad (2)$$

Here $\sigma_y$ represents the stress that arises as a consequence of the progressive yield process, and $\sigma_h$ the stress caused by the dislocation strain hardening within crystal grains.

Using Figure 3(b) as an illustration for the derivation of stress $\sigma_y(\varepsilon)$ from the Yield Volume Fraction $\beta(\varepsilon)$, we note that when the sample strain increases from $e$ to $e+de$, the *additional* yielding volume fraction in this range is $\beta'(e)de$. When the total strain reaches a value $\varepsilon$ that is less than $\varepsilon_1$, the polycrystal behaves elastically, and hence, the stress can be calculated by Hooke's law $\sigma(\varepsilon) = E\varepsilon$. After yielding has occurred, it causes a reduction in the load-bearing capacity of the material in the plastic region. The stress reduction caused by the *additional* yielding volume fraction $\beta'(e)de$ between $e$ and $e+de$ is equal to $E(\varepsilon - e)\beta'(e)de$. The stress $\sigma_y$ in the yield process can be calculated by the integration:

$$\sigma_y(\varepsilon) = E\varepsilon - E\int_0^\varepsilon (\varepsilon - e)\beta'(e)de \qquad (3)$$

Integrating by parts gives:

$$\sigma_y(\varepsilon) = E\varepsilon - ((\varepsilon - \varepsilon)\beta(\varepsilon) - \varepsilon\beta(0)) - E\int_0^\varepsilon \beta(\varepsilon)\,de = E\varepsilon - E\int_0^\varepsilon \beta(\varepsilon)\,de \qquad (4)$$

Differentiating both sides with respect to $\varepsilon$ yields the following expression:

$$\beta(\varepsilon) = 1 - \frac{1}{E}\frac{d\sigma_y}{d\varepsilon} \qquad (5)$$

Equation (4) expresses the current stress as a function of the Yield Volume Fraction $\beta(\varepsilon)$, whilst Equation (5) represents the inversion of this relationship that expresses $\beta(\varepsilon)$ in terms of $\sigma_y(\varepsilon)$. The additional hardening stress $\sigma_h$ can be approximated by a simple linear hardening law:

$$\sigma_h(\varepsilon) = E_m\varepsilon\beta(\varepsilon), \qquad (6)$$

Here $E_m$ is the tangent modulus, and the multiplier $\beta(\varepsilon)$ reflects the fraction of the total volume that contributes to additional hardening.

Once CPFE simulation of deformation is carried out, the cumulative yield volume fraction $\beta(\varepsilon)$ can be obtained from results analysis. It is found (Figures 4 and 5) that the cumulative Yield Volume Fraction $\beta(\varepsilon)$ can be well described by the cumulative distribution function (CDF) of lognormal or Gumbel extreme value distribution, and the derivative of $\beta(\varepsilon)$ (the incremental Yield Volume Fraction) is in good agreement with the probability density function (PDF) of lognormal or Gumbel distribution. The Yield Volume Fraction method can be developed further based on the above finding. The stress-strain curves function $\sigma(\varepsilon)$ can be then determined from Equation (5) and Equation (6) by substitution into Equation (2).

The probability density function *f(x)* and cumulative distribution function *F(x)* of the lognormal distribution are expressed as:

$$f(x|M,S) = \frac{A_1}{xS\sqrt{2\pi}} \exp\left\{\frac{-(\ln(x)-M)^2}{2S^2}\right\} \quad (7)$$

$$F(x|M,S) = A_1 \Phi\left(\frac{\ln(x)-M}{S}\right) \quad (8)$$

where $M$ and $S$ are respectively the mean and standard deviation of the variable $\ln(x)$, $\Phi$ is the CDF of the standard normal distribution.

The probability density function *g(x)* and cumulative distribution function *G(x)* of the Gumbel distribution are given by:

$$g(x|\mu,\sigma) = A_2 \exp\left[-\exp\left[-\frac{(x-\mu)}{\sigma}\right] - \frac{(x-\mu)}{\sigma}\right] \quad (9)$$

$$G(x|\mu,\sigma) = A_2 \int \exp\left[-\exp\left[-\frac{(x-\mu)}{\sigma}\right] - \frac{(x-\mu)}{\sigma}\right] dx \quad (10)$$

where $\mu$ and $\sigma$ are the location parameter and scale parameter, respectively.

## 2.4 The Ramberg-Osgood relationship and its connection to the proposed method

The Ramberg-Osgood relationship shown in Equation (1) is a simple method to reproduce experimental stress-strain curves. First, Youngs' modulus $E$ can be determined from the slope of the elastic region in strain-stress curves. The parameters $K$ and $n$ are then can be fitted to the following power-law relationship:

$$\varepsilon_{pl} = \varepsilon - \frac{\sigma}{E} = \left(\frac{\sigma}{K}\right)^n \quad (11)$$

The incremental form of the Ramberg-Osgood relationship can be calculated as follows:

$$\frac{d\varepsilon}{d\sigma} = \frac{1}{E} + \frac{n\sigma^{n-1}}{K^n} \quad (12)$$

Therefore

$$\frac{d\sigma}{d\varepsilon} = \frac{1}{\frac{1}{E} + \frac{n\sigma^{n-1}}{K^n}} \tag{13}$$

According to Equation (5), the cumulative yield volume fraction function $\beta(\varepsilon)$ for the Ramberg-Osgood relationship is obtained:

$$\beta(\varepsilon) = 1 - \frac{1}{\left(1 + \frac{nE\sigma^{n-1}}{K^n}\right)} \tag{14}$$

The incremental Yield Volume Fraction $\gamma(\varepsilon)$ is obtained as the derivative of the Yield Volume Fraction function $\beta(\varepsilon)$ with respect to strain $\varepsilon$ for the Ramberg-Osgood relationship in the form:

$$\frac{d^2\sigma}{d\varepsilon^2} = \frac{d}{d\varepsilon}\left(\frac{d\sigma}{d\varepsilon}\right) = \frac{d}{d\sigma}\left(\frac{d\sigma}{d\varepsilon}\right) \cdot \frac{d\sigma}{d\varepsilon} = \frac{-(n-1)n\sigma^{n-2}}{K^n\left(\frac{1}{E} + \frac{n\sigma^{n-1}}{K^n}\right)^2} \cdot \frac{1}{\frac{1}{E} + \frac{n\sigma^{n-1}}{K^n}} \tag{15}$$

$$\gamma(\varepsilon) = \frac{d\beta(\varepsilon)}{d\varepsilon} = -\frac{1}{E}\frac{d^2\sigma}{d\varepsilon^2} = \frac{(n-1)n\sigma^{n-2}}{E \cdot K^n\left(\frac{1}{E} + \frac{n\sigma^{n-1}}{K^n}\right)^3} \tag{16}$$

When once the parameters *K* and *n* are determined by fitting the stress-strain curve, the cumulative and derivative of the Yield Volume Fraction for the Ramberg-Osgood relationship can be then calculated by Equations (14) and (16), respectively.

## 3. Results and discussion

In the current study, the following procedure was followed to calibrate the fitting parameters for the Ramberg-Osgood relationship, lognormal and Gumbel distribution:

(1) CPFE simulation was employed to fit the experimental stress-strain curves shown in Figure 1. The crystal plasticity model parameters were optimized by fitting the simulation results to macroscopic experimental stress-strain curves and experimental neutron diffraction measurements, as described in the reference (Chen et al., 2022).

(2) Once the best-fit parameters were found for CPFE simulation, the cumulative Yield Volume Fraction $\beta(\varepsilon)$ and the incremental Yield Volume Fraction $\gamma(\varepsilon)$ could be readily obtained from CPFE simulation results.

(3) According to Equations (7), (9) and (16), the PDF for lognormal distribution, Gumbel distribution and Ramberg-Osgood relationship were used to fit the incremental Yield Volume Fraction $\gamma(\varepsilon)$ obtained in CPFE simulation. Least squares optimization was employed to find the best fitting parameters.

(4) Once the best fitting parameters, i.e. $M$, $S$ and $A_1$ for lognormal distribution; $\mu$, $\sigma$ and $A_2$ for Gumbel distribution; $K$ and $n$ for the Ramberg-Osgood relationship were determined, and the cumulative yield volume fraction for these three types of methods can be calculated by Equations (8), (10) and (14).

(5) Finally, the stress-strain curves reproduced by a lognormal distribution, Gumbel distribution and R-O relationship can be obtained according to Equations (2), (5) and (6).

The best-fitting parameters for lognormal distribution, Gumbel distribution and R-O relationship are listed in Table 1. The tangent modulus for developed plastic flow regime $E_m$ is determined to be 3124 MPa. Figure 4 and Figure 5 compare the cumulative and incremental Yield Volume Fraction obtained from CPFE simulation, R-O relationship and the presently introduced YVF method. The cumulative and incremental Yield Volume Fraction prediction obtained from lognormal and Gumbel distribution are found to be in excellent agreement with the CPFE simulation result, while a significant difference is observed for the R-O relationship. The R-O relationship is capable of predicting the cumulative and derivative yield volume fraction when the total strain is less than 0.23%. However, it significantly underestimates the rate and the cumulative volume of the yield process during further loading above 0.23% strain. A greater volume of material behaves elastically according to the R-O equation, and consequently, R-O tends to overestimate the stress-strain curves and the deviation becomes larger as the total strain increases. The predictions of the proposed Yield Volume Fraction method using both lognormal and Gumbel distribution show significantly better agreement with the CPFE prediction and experimental curves, especially in the elastic-plastic transition regime. The above comparison reveals the advantage of the proposed method: it can reproduce the experimental stress-strain curves (especially the 'knee' region of the curves) with sufficient accuracy as it employs a statistical method to describe the yield volume fraction in a progressive yield process.

Table 1. Material parameters used in R-O relationship and the proposed method.

| Methods | R-O relationship | | | Lognormal distribution | | | Gumbel distribution | | |
|---|---|---|---|---|---|---|---|---|---|
| Parameters | $E$(MPa) | $K$ | $n$ | $A_1$ | $M$ | $S$ | $A_2$ | $\mu$ | $\sigma$ |
| Value | 220000 | 1100 | 11.5 | 0.956 | 0.00255 | 0.23 | 1860.5 | 0.00237 | 0.000524 |

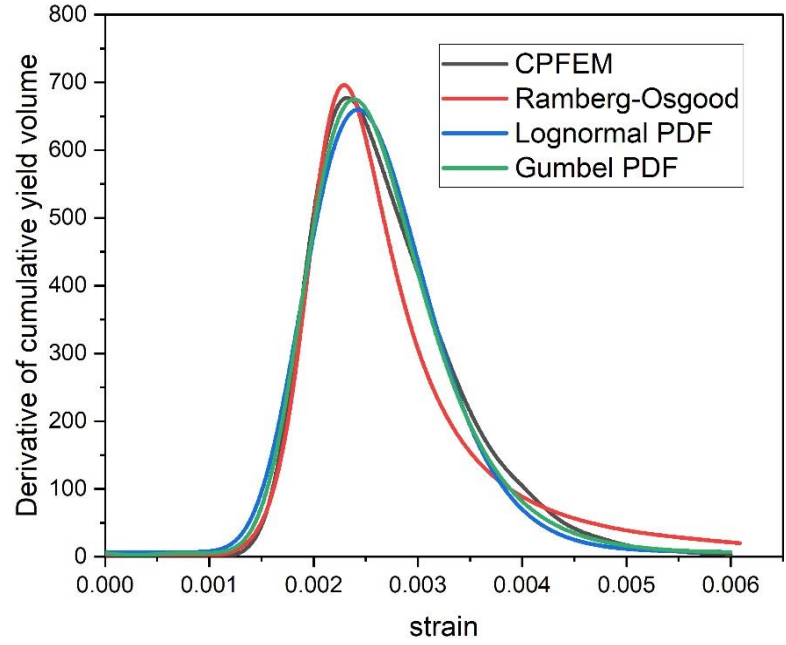

**Figure 4**. The derivative of yield volume fraction obtained from CPFE simulation, R-O relationship and the proposed method.

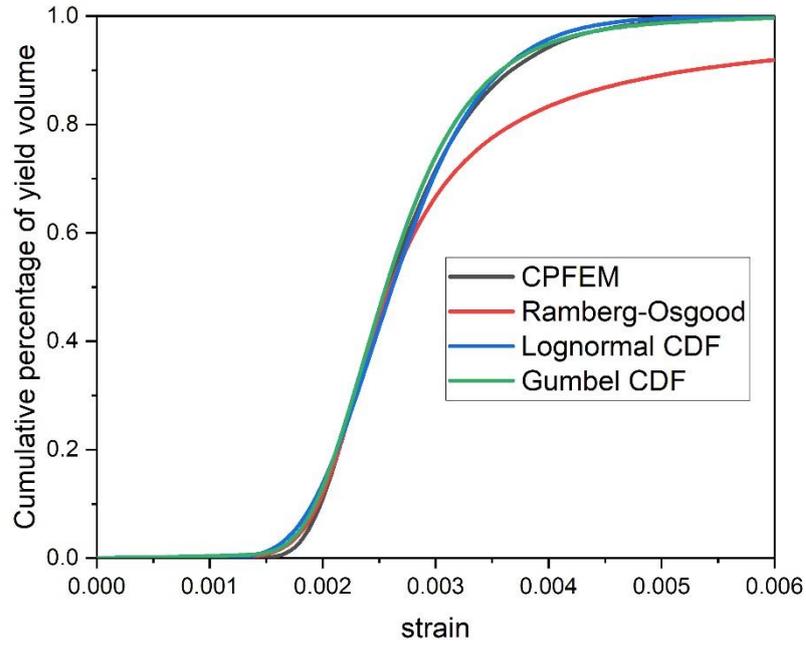

**Figure 5**. The cumulative yield volume fraction obtained from CPFE simulation, R-O relationship and the proposed method.

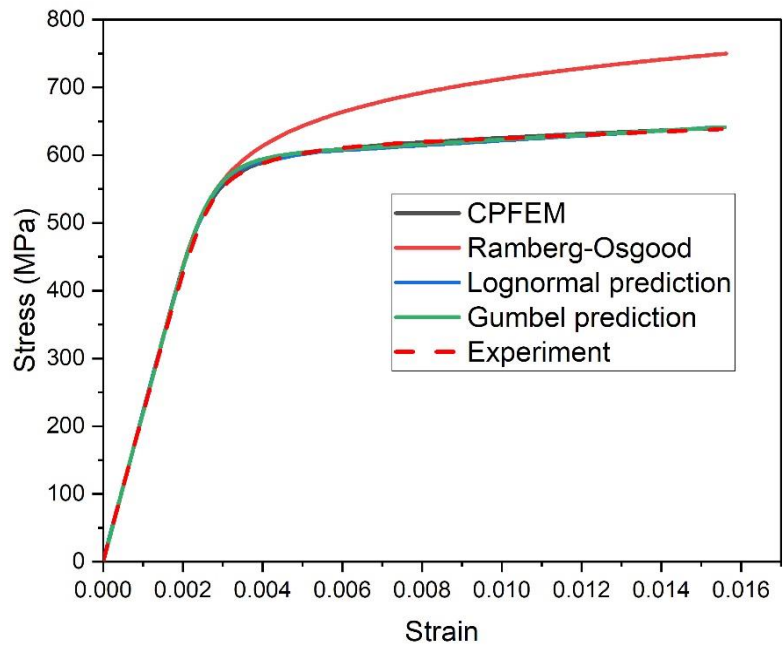

**Figure 6**. Stress-strain curves obtained from experiment, CPFE simulation, R-O relationship and the proposed YVF method.

## 4. Conclusions

It has been well accepted that the Ramberg-Osgood relation can accurately reproduce stress-strain curves of metallic materials up to 0.2% yield stress but is not suitable for higher ranges of stresses. In the present study, the limitation of the Ramberg-Osgood equation was analysed by comparing with the yield volume fraction integrated method. R-O tends to underestimate the incremental and cumulative Yield Volume Fraction when the strain level is above 0.23%, and hence overpredict the stress-strain curves in the 'knee' region of the stress-strain curve.

In the present paper, a new method was proposed for describing the stress-strain behaviour using the Yield Volume Fraction function. The relationship between stress-strain curves and the Yield Volume Fraction function was derived. Lognormal and Gumbel distribution could provide excellent agreement with the Yield Volume Fraction obtained from the experiment and CPFE simulation. The accuracy of the proposed model was demonstrated by comparing its prediction with CPFE simulation, the Ramberg-Osgood model and experimental observations. It was found that YVF descriptions obtained from lognormal and Gumbel distribution possess the advantage of improved accuracy, especially in the elastic-plastic transition regime.

## CRediT authorship contribution statement

**Jingwei Chen:** Conceptualization, Formal analysis, Validation, Visualization, Writing – original draft. **Alexander M. Korsunsky:** Conceptualization, Methodology, Supervision, Formal analysis, Writing – review & editing.

## Acknowledgements

This research did not receive any specific grant from funding agencies in the public,

commercial, or not-for-profit sectors.

## Data availability statement

The datasets used and/or analysed during the current study are available from the corresponding author on reasonable request.